\documentclass[conference]{IEEEtran}
\IEEEoverridecommandlockouts

\usepackage{cite}
\usepackage{amsmath,amssymb,amsfonts}
\usepackage{algorithmic}
\usepackage{graphicx}
\usepackage{textcomp}
\usepackage{xcolor}
\usepackage{stfloats}
\def\BibTeX{{\rm B\kern-.05em{\sc i\kern-.025em b}\kern-.08em
    T\kern-.1667em\lower.7ex\hbox{E}\kern-.125emX}}
\begin{document}

\title{Weight-Guided Constraints for Body Model and Lead Selection in Pediatric CIED MRI Safety Simulations\\

\thanks{This work was supported under NIH Grants R01EB036272, R01HL168859, and R01EB034377.}
}

\author{
\IEEEauthorblockN{
Safa Hameed, BS\IEEEauthorrefmark{1}\IEEEauthorrefmark{2},
Kaylee Henry, PhD\IEEEauthorrefmark{1},
Fuchang Jiang, MS\IEEEauthorrefmark{1}\IEEEauthorrefmark{2},
Bhumi Bhusal, PhD\IEEEauthorrefmark{2},
Halley Dillenbeck\IEEEauthorrefmark{3},\\
Lindsey Gakenheimer-Smith, MD, MScPH\IEEEauthorrefmark{3},\
Gregory Webster, MD, MPH\IEEEauthorrefmark{3},\
Laleh Golestanirad, PhD\IEEEauthorrefmark{1}\IEEEauthorrefmark{2}
}
\IEEEauthorblockA{\IEEEauthorrefmark{1}Department of Biomedical Engineering, Northwestern University, Evanston, IL, USA}
\IEEEauthorblockA{\IEEEauthorrefmark{2}Department of Radiology, Feinberg School of Medicine, Northwestern University, Chicago, IL, USA}
\IEEEauthorblockA{\IEEEauthorrefmark{3}Division of Cardiology, Department of Pediatrics, Ann \& Robert H. Lurie Children's Hospital, Chicago, IL, USA}
\IEEEauthorblockA{Corresponding author: Laleh Golestanirad (laleh.rad1@northwestern.edu)}
}

\maketitle

\begin{abstract}
Pediatric patients with cardiac implantable electronic devices (CIEDs) face limited MRI access due to RF-induced heating, and computational modeling is increasingly used to characterize this risk. The validity of these simulations, however, depends on pairing body models with clinically realistic lead configurations, guidance that is currently lacking. We retrospectively analyzed 302 CIED surgeries in 281 pediatric patients to derive weight-based constraints for simulation design. Weight alone discriminated epicardial from endocardial lead implantation with AUC = 0.90, and adding age and height yielded no improvement, supporting weight as a sufficient single-parameter selection metric. The probabilistic crossover between approaches occurred at 44~kg, substantially higher than the 10 to 15~kg threshold commonly cited in the literature, with a broad transition zone of 21 to 66~kg in which both lead types were routinely used. Lead length was likewise weight-constrained: only 25~cm leads were observed in patients below 6~kg, and leads of 45~cm or longer were uncommon below 50~kg. These findings yield a three-tier framework, with epicardial-only configurations below 21~kg, dual configurations within 21 to 66~kg, and weight-thresholded lead lengths throughout, enabling MRI safety simulations to focus on clinically realizable anatomy and device combinations.
\end{abstract}

\section{Introduction}
Pediatric patients with cardiac implantable electronic devices (CIEDs) face significant barriers to MRI access due to the risk of radiofrequency (RF)-induced heating \cite{b1,b2}. This risk is driven by the coupling between the MRI transmit field and the implanted lead, a phenomenon highly sensitive to the lead’s trajectory and routing geometry \cite{b3, b4, b5, b6, b7, b8}. In pediatric practice, these geometries vary dramatically based on patient size. Surgeons implant epicardial systems, where leads are sutured to the epicardium and connected to an abdominal generator, or electrophysiologists implant endocardial systems, which are routed transvenously to a generator affixed to the pre-pectoral fascia. Because these configurations possess fundamentally different trajectories, they exhibit distinct RF coupling mechanisms \cite{b3,b7, b9, b10, b11}.

While MR-conditional labeling exists for many adult endocardial systems, no epicardial systems have received MR-conditional labeling, limiting pediatric access to MRI \cite{b12}. This regulatory gap forces clinicians to rely on alternative imaging modalities that involve ionizing radiation and may provide suboptimal diagnostic capability \cite{b13, b14}.

Computational electromagnetic modeling is increasingly used to address this gap, allowing researchers to quantify RF heating across device configurations that cannot be exhaustively tested experimentally \cite{b15, b16}. However, the validity of these simulations depends critically on selecting body models that accurately reflect the patient population for the device being studied. Currently, there is no quantitative guidance linking patient size to lead type or lead length selection. Without such data, researchers risk overlooking clinically important edge cases, such as small infants with unusual systems, or simulating unrealistic scenarios, such as endocardial leads in infants, that rarely occur in clinical practice. Given the high computational cost of these studies, data-driven constraints are needed to focus simulations on clinically relevant anatomy--device combinations \cite{b17}.

In this retrospective study of surgical records, we correlated patient size with lead type and length. We converted these findings into simple, weight-based parameters for MRI safety research. These parameters will help engineers select clinically plausible combinations of body models and devices, ensuring their simulations reflect real-world patients.

\begin{figure*}[!b]
  \centering
  \includegraphics[width=\textwidth]{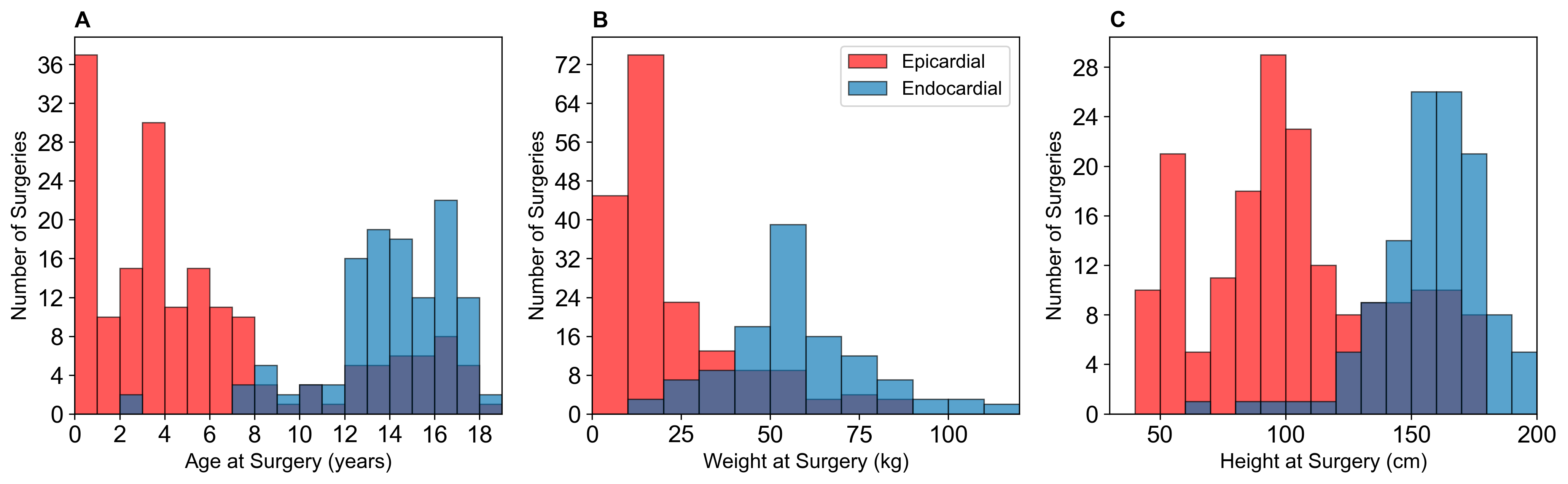}
  \caption{Pediatric cohort anthropometrics stratified by lead type (surgery-level). Histograms show distributions of (A) age, (B) weight, and (C) height at the time of surgery for epicardial (n = 183) versus endocardial CIED (n = 119) implants. Epicardial implants cluster in younger, smaller patients, while endocardial implants occur predominantly in larger adolescents, motivating size-guided cohort definitions for simulation studies.}
  \label{fig:wide}
\end{figure*}
\section{Method}

\subsection{Data Source}\label{AA}
We performed a retrospective analysis of pediatric CIED implantations at Ann \& Robert H.\ Lurie Children's Hospital of Chicago using data from surgical, electrophysiology, and electronic medical records, including a Medtronic device database (``PaceArt''). We linked device-specific parameters with patient anthropometric data at the time of implantation. All analyses were performed on de-identified registry extracts. Patient demographics are summarized in Table~I, and distributions of age, weight, and height by approach are shown in Fig.~1.

\subsection{Study Population}
The length of each epicardial lead is selected by the cardiovascular surgeon at the time of surgery based on clinical judgment that incorporates patient size, the route from the myocardium to the implantable pulse generator (IPG), and practical hospital inventory. Therefore, each surgical procedure was treated as an independent data point. Patients who underwent multiple operations (e.g., system upgrades) contributed multiple observations; results were confirmed in a sensitivity analysis restricted to the first surgery per patient.
Inclusion criteria were: (1) age $\leq$ 18 years, (2) documented lead implantation, and (3) complete height and weight data.

\subsection{Variable Definition}
We extracted patient age, weight, height, and lead parameters (model and length) for each procedure. Leads were classified by model number as epicardial (Medtronic 4965, 4968, and 5071) or endocardial (Medtronic 3830, 5076, 6937A, 6935, 5086, and 6947M), and both pacing and high-voltage (defibrillation) leads were included. Coded registry values were mapped to physical lead lengths (e.g., 25, 35, 45, 52, and 55~cm).

To characterize size-dependent selection patterns, we also analyzed a lead-level dataset in which each implanted lead was treated as an independent observation.
\begin{table}[t]
\caption{Patient Demographics by Lead Type}
\label{tab:demo}
\centering
\renewcommand{\arraystretch}{1.2}
\setlength{\tabcolsep}{3pt}
\footnotesize

\begin{tabular}{|c|c|c|c|c|}
\hline
 & \textbf{\textit{\shortstack{All\\($n=302$)}}}
 & \textbf{\textit{\shortstack{Epicardial\\($n=183$)}}}
 & \textbf{\textit{\shortstack{Endocardial\\($n=119$)}}}
 & \textbf{\textit{$p$}} \\
\hline

\multicolumn{4}{|c|}{\textbf{Age at implant, years}} & \\ \hline
\textbf{Mean $\pm$ SD}   & $9 \pm 6$      & $6 \pm 5$        & $14 \pm 3$       & $< 0.001$ \\ \hline
\textbf{Median (IQR)}    & \shortstack{9\\(4--15)}      & \shortstack{3.9\\(1--8)}       & \shortstack{14\\(13--16)}      &  \\ \hline
\textbf{Range}           & 0--18          & 0--18            & 2--18            &  \\ \hline

\multicolumn{4}{|c|}{\textbf{Weight, kg}} & \\ \hline
\textbf{Mean $\pm$ SD}   & $36.0 \pm 26.0$ & $21.9 \pm 18.7$   & $57.7 \pm 19.9$   & $< 0.001$ \\ \hline
\textbf{Median (IQR)}    & \shortstack{29.6\\(13.9--55.3)} & \shortstack{15.3\\(10.0--28.2)} & \shortstack{55.7\\(46.3--68.7)} &  \\ \hline
\textbf{Range}           & 1.6--117.5      & 1.6--89.6         & 13.0--117.5       &  \\ \hline

\multicolumn{4}{|c|}{\textbf{Height, cm}} & \\ \hline
\textbf{Mean $\pm$ SD}   & $125.2 \pm 40.8$ & $104.1 \pm 36.4$  & $157.7 \pm 21.2$  & $< 0.001$ \\ \hline
\textbf{Median (IQR)}    & \shortstack{133.2\\(95.2--161.0)} & \shortstack{100.0\\(78.7--129.1)} & \shortstack{160.0\\(148.1--170.5)} &  \\ \hline
\textbf{Range}           & 41.4--197.4      & 41.4--176.5       & 65.0--197.4        &  \\ \hline

\end{tabular}
\end{table}

\subsection{Statistical Analysis}
Demographics were summarized using mean $\pm$ standard deviation, as well as median [IQR] (and range where appropriate). Comparisons between epicardial and endocardial groups were performed using Mann--Whitney U tests for continuous variables. Univariate logistic regression was used to evaluate age, weight, and height as predictors of epicardial approach. Discrimination was quantified using the area under the receiver operating characteristic curve (AUC) with 95\% bootstrap confidence intervals (1000 iterations). Because pediatric anthropometric variables are highly collinear, we used univariate models to identify the most practical single-variable discriminator for cohort definition and evaluated a multivariable model (age, height, and weight) as a sensitivity analysis. We report a descriptive transition point defined as the predictor value at which the fitted probability of epicardial approach equals 0.50.

To provide body-model selection guidance, we computed lead-type-specific weight coverage ranges using the 5th--95th percentiles (covering 90\% of procedures). Lead length distributions were tabulated across weight categories to identify weight ranges in which specific lead lengths were absent and to translate these constraints into simulation design guidance. Weight categories were selected based on inspection of lead length distributions to identify thresholds at which predominant lead lengths shifted. This process yielded five categories: $< 6$~kg, 6--13~kg, 13--21~kg, 21--48~kg, and $> 48$~kg. These categories provide additional resolution within the $<10$--15~kg population, where transvenous access may be limited by venous constraints and where some centers reserve endocardial leads for larger children (e.g., $> 15$--20~kg). All analyses were performed in Python 3.12 using SciPy, scikit-learn, and pandas, with $\alpha = 0.05$.

\section{Results}
\subsection{Study Population}
\label{study_pop}
A total of 305 pediatric CIED surgeries were identified. After excluding procedures with missing height or weight data at the time of implantation ($n = 3$), the final surgery-level cohort comprised 302 surgeries. Epicardial implantation accounted for 183/302 (60.6\%) of surgeries, whereas endocardial implantation accounted for 119/302 (39.4\%).

Breaking down the study population by implant approach, patients receiving epicardial leads showed a primary cluster in early childhood (age $< 5$ years; 56.3\%) and a secondary cluster in adolescence (age $\geq 10$ years; 21.9\%), with relatively few implantations between ages 8 and 10 years (2.2\%).

Patients receiving epicardial leads were significantly younger (median 3.9 vs.\ 14 years, $p < 0.001$), weighed less (median 15.3 vs.\ 55.7~kg, $p < 0.001$), and were shorter (median 100 vs.\ 160~cm, $p < 0.001$) than those receiving endocardial leads (Table~I). The difference in weight between groups was substantial, with a large effect size (Cohen's $d = 1.76$).

\subsection{Patient size predicts lead type }\label{size_pred_lead_type}
Among the univariate predictors, weight yielded the highest discrimination for epicardial lead implantation (AUC = 0.90, 95\% CI: 0.86--0.94), marginally outperforming age (AUC = 0.87, 95\% CI: 0.82--0.91) and height (AUC = 0.88, 95\% CI: 0.84--0.92). A sensitivity analysis restricted to the first surgery per patient yielded similar discrimination (AUC = 0.91).

Importantly, a multivariable model combining age, height, and weight did not improve discrimination relative to weight alone (AUC = 0.90, 95\% CI: 0.86--0.93), supporting the use of weight as a sufficient single-parameter metric for body model selection. The probabilistic crossover point, defined as the weight at which the predicted probability of epicardial implantation equals 50\%, occurred at approximately 44~kg (Fig.~2). However, this metric reflects only the central tendency; clinical practice exhibited a broad transition zone around this threshold.

\begin{figure}[!t]
  \centering
  \includegraphics[width=\columnwidth]{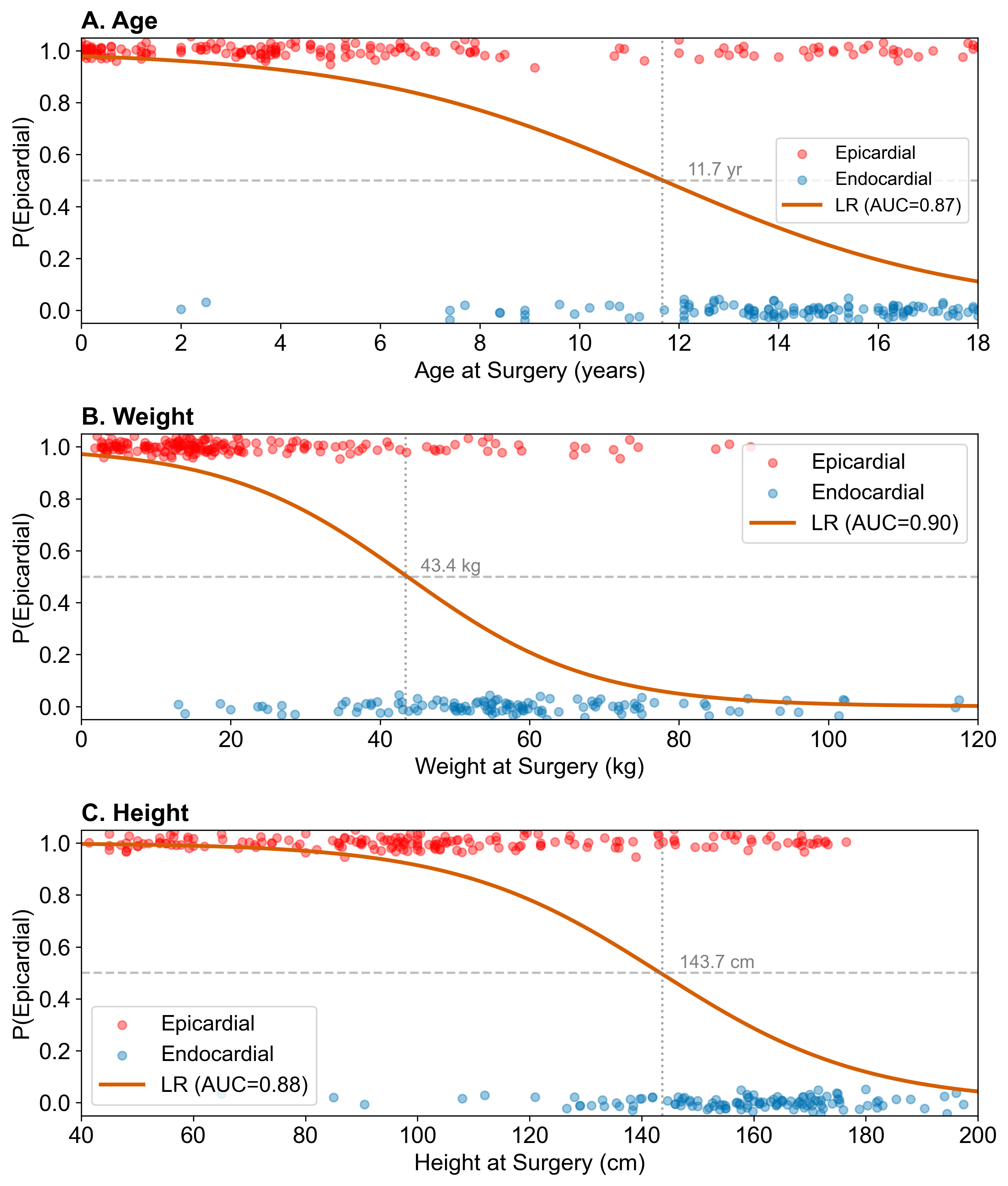}
  \caption{Weight best discriminates epicardial versus endocardial approach for defining clinically representative simulation cohorts. Surgery-level data are shown with univariate logistic fits for (A) age, (B) weight, and (C) height. Model discrimination is quantified by AUC (weight: 0.90; height: 0.88; age: 0.87). The vertical dashed lines mark the predictor value at which the fitted probability of epicardial implantation equals 0.50 (descriptive transition point).}
  \label{fig:fig2}
\end{figure}

\subsection{Simulation guidance: weight ranges and lead-length}\label{sim_design_guidance}
To translate these findings into robust body-model selection guidance, we computed lead-type-specific weight coverage ranges using the 5th--95th percentiles. Epicardial implants were prevalent in patients weighing 3--66~kg, whereas endocardial implants were prevalent in patients weighing 21--96~kg. The substantial overlap between these ranges (21--66~kg) defines a transition zone in which simulations of both lead types are required to ensure comprehensive safety coverage.

To characterize lead length selection patterns, we additionally analyzed 509 implanted leads from 281 unique patients and found that lead length was also strongly constrained by patient weight in this cohort. In the smallest patients ($< 6$~kg), only 25~cm epicardial leads were observed. In the 6--13~kg range, 25~cm epicardial leads remained predominant, with limited use of 35~cm leads. Longer leads were not observed in patients weighing 13--21~kg and only appeared at higher weights, where both lead types and multiple lead lengths were prevalent (Fig.~3). These findings define lower weight thresholds below which specific lead lengths were absent in this cohort, providing practical guidance for constructing clinically representative simulation matrices.

\section{Discussion}
This study quantifies how pediatric CIED implant approach varies with patient size and translates these clinical patterns into actionable constraints for MRI safety study design. Our results identify weight as the most practical single-parameter discriminator for epicardial versus endocardial lead selection (AUC = 0.90). Notably, adding age and height to the predictive model yielded negligible improvement, confirming that weight, a standard metric in virtual body model libraries, is sufficient for defining simulation cohorts.

\subsection{Clinical Validity and the Transition Zone}
\label{SCM}
The strong size dependence observed in this cohort aligns with general principles of pediatric CIED implantation, in which endocardial leads are often deferred in smaller children to preserve venous patency. There is no universal threshold for transitioning from epicardial to transvenous systems. However, despite reports of transvenous pacing systems in patients weighing $< 10$~kg \cite{b18, b19}, most authors describe a transition zone with a lower boundary of 10--15~kg for pacing leads \cite{b20}.

Some centers routinely defer transvenous implantation until children reach larger body sizes, even for pacing systems, and there is broad consensus that transvenous ICD systems require greater venous caliber and body habitus than pacing systems \cite{b21}. However, despite smaller crossover sizes reported in the literature, our data from a single high-volume center revealed a probabilistic crossover point at approximately 44~kg and a broad transition zone spanning 21--66~kg, within which both epicardial and endocardial approaches were frequently used.

This discrepancy highlights that, although endocardial pacing is technically feasible in small children, it is not consistently practiced until patients are substantially larger. This pattern may reflect body size considerations, such as venous caliber and concerns regarding long-term venous occlusion, as well as intracardiac anatomy or congenital abnormalities that preclude transvenous access. For the simulation engineer, the specific clinical rationale is less important than the central observation that a wide transition zone exists. Assuming a binary switch to endocardial systems at 15~kg would prematurely exclude epicardial models for many school-aged children and fail to capture a common clinical reality.
\begin{figure*}[!t]
  \centering
  \includegraphics[width=\textwidth]{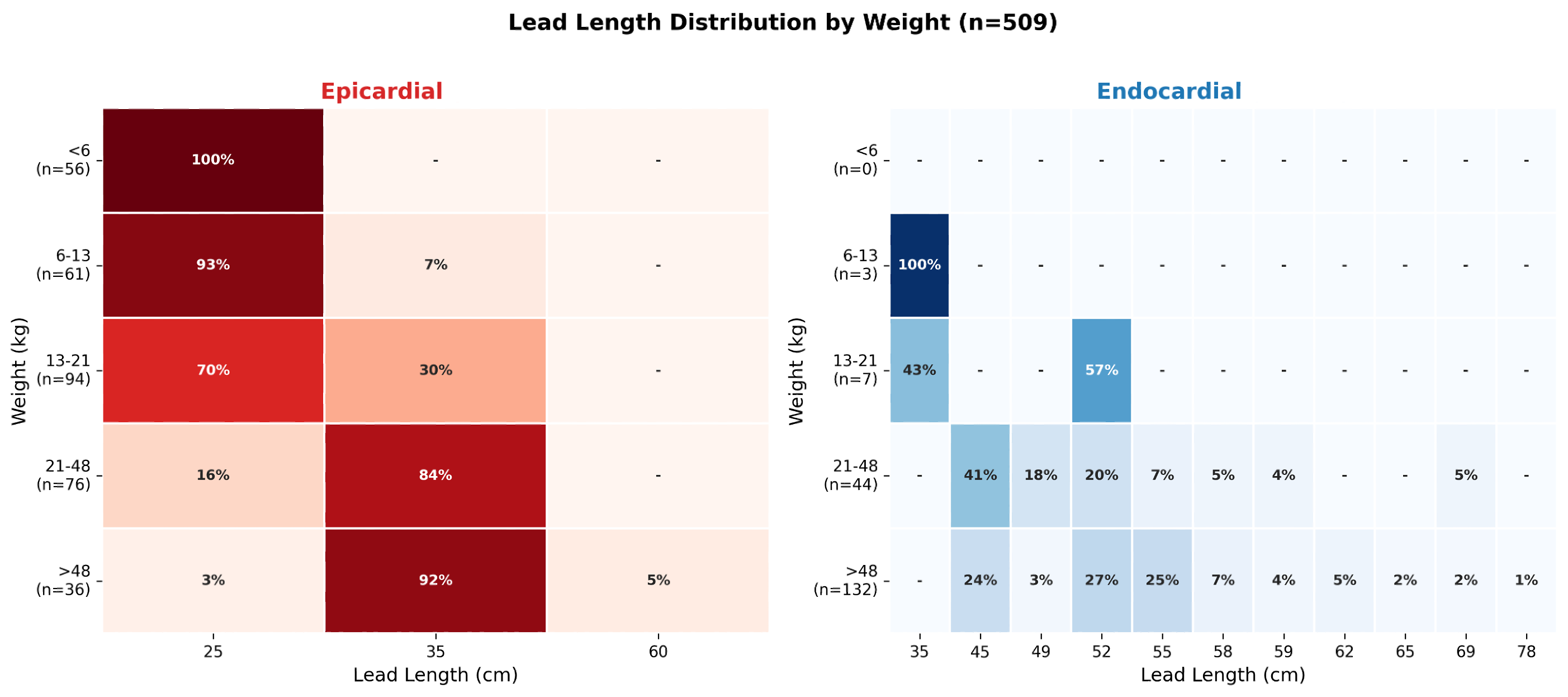}
  \caption{ Lead-length distribution by weight category and lead type (instance-level, n = 509). Cells show percentage of leads within each weight bin. Epicardial leads are predominantly 25-35 cm, while endocardial leads are predominantly 45-55 cm above 21 kg.}
  \label{fig:wide}
\end{figure*}

\subsection{Data-Driven Constraints for Simulation Design}
Current computational safety studies typically rely on case-by-case pairings of body models and devices, and quantitative guidance is not available to ensure that simulated anatomy and device combinations reflect clinical reality. As pediatric MRI safety simulations expand toward larger, more systematic cohorts, the risk of including clinically unlikely scenarios, such as long endocardial leads in infants, becomes a practical concern that can inflate computational cost without improving clinical relevance. To address this gap, we propose a weight-stratified simulation framework based on the 90\% coverage intervals identified in this study:
\begin{itemize}
\item Small Pediatric ($<21$ kg): Simulations should prioritize epicardial configurations. Endocardial leads were statistically rare in this weight range, with a 5th percentile threshold of 21~kg, and may be excluded unless the goal is to investigate specific outlier scenarios.
\item Transition Cohort (21–66 kg): This range represents the overlap zone, where clinical practice varies most. Safety assessments for body models in this weight class should include both epicardial and endocardial configurations to ensure comprehensive coverage.
\item Lead Length Constraints: We identified strict lower bounds for lead selection as a function of patient weight. For example, 25~cm leads were the exclusive option in patients weighing $< 6$~kg, and leads $\geq 45$~cm were not commonly used until patients exceeded 50~kg. These practical constraints allow researchers to eliminate unrealistic lead-length combinations from simulation matrices.
\end{itemize}

\subsection{Limitations and Future Directions}
This analysis reflects a single-center cohort using Medtronic lead models; consequently, the absolute prevalence thresholds reported here may vary across institutions with different surgical practices or device portfolios. In addition, our primary analysis was performed at the procedure level, although sensitivity analyses indicated that repeated observations from the same patient did not materially affect the primary predictors.

Finally, because some patients are paced using high-voltage leads and many ICD systems include a separate atrial pacing lead, we did not analyze pacing and high-voltage leads separately. Distinguishing these lead categories may reveal additional nuances relevant to MRI safety modeling and should be considered in future work. Extending this framework to multicenter registries would improve the generalizability of these constraints and further refine the practical lookup tables needed to standardize pediatric MRI safety simulations.

\section{Conclusion}
This study establishes weight as the primary anthropometric metric for defining pediatric CIED simulation cohorts. Although clinical practice demonstrates clear preferences for epicardial systems in infants and endocardial systems in adolescents, we identified a broad transition zone spanning 21--66~kg in which both configurations should be modeled to ensure comprehensive safety coverage. In addition, observed lead-length selection patterns impose practical constraints in the smallest patients, eliminating the need to simulate long leads in infant body models. By adhering to these data-driven boundaries, MRI safety researchers can construct simulation matrices that are both computationally efficient and clinically robust, ensuring that virtual risk assessments reflect real-world pediatric patient populations.

\end{document}